\documentclass[conference]{IEEEtran}
\usepackage{amssymb,amsmath}
\usepackage[dvips]{graphicx}
\usepackage{color}
\usepackage[dvips]{graphicx}
\newcounter{proposition}

\newcommand{\nothing}[1]{}

\newcommand{\beq}[1]{\begin{equation}\label{#1}}
	\newcommand{\eeq}{\end{equation}}
\newcommand{\req}[1]{(\ref{#1})}
\newcommand{\bmu}[1]{\begin{multline}\label{#1}}
	\newcommand{\emu}{\end{multline}}

\newcommand{\eq}{\triangleq}

\newcommand{\x}{{\textbf{\textit{x}}}}

\renewcommand{\u}{{\bf u}}
\renewcommand{\v}{{\bf v}}
\renewcommand{\S}{{\cal S}}

\renewcommand{\S}{{\mathcal{S}}}

\renewcommand{\chi}{\upsilon}

\newcommand{\lev}{\left\lceil}
\newcommand{\riv}{\right\rceil}
\renewcommand{\le}{\leqslant}

\renewcommand{\leq}{\leqslant}
\renewcommand{\geq}{\geqslant}

\title{On a  Hypergraph Approach to Multistage
   Group Testing  Problems}
\author{\IEEEauthorblockN{ A. G. D'yachkov, I.V. Vorobyev, N.A. Polyanskii and V.Yu. Shchukin}
	\IEEEauthorblockA{Lomonosov Moscow State University,
		Moscow, Russia\\
		Email: agd-msu@yandex.ru,\quad vorobyev.i.v@yandex.ru,\quad nikitapolyansky@gmail.com,\quad
		vpike@mail.ru}}
\begin{document}
\maketitle
\begin{abstract}
Group testing is a well known search problem that consists in detecting up to $s$
defective elements of the set $[t]=\{1,\ldots,t\}$ by carrying out tests on properly chosen subsets
 of $[t]$. In classical group testing the goal is to find all defective elements by
using the minimal possible number of tests. In this paper we consider multistage group testing. We propose a general idea how to use a hypergraph approach to searching defects. For the case $s=2$, we design an explicit construction, which makes use of $2\log_2t(1+o(1))$ tests in the worst case  and consists of $4$ stages. For the general case $s>2$, we provide an explicit construction, which uses $(2s-1)\log_2t(1+o(1))$ tests and consists of $2s-1$ rounds.
\end{abstract}	

\section{Introduction}\label{Introd}

Group testing is a  very natural combinatorial problem that consists in detecting up to $s$
defective elements of the set of objects $[t]=\{1,\ldots,t\}$ by carrying out tests on properly chosen subsets (pools)
 of $[t]$. The test outcome is positive if the tested pool contains one
or more defective elements; otherwise, it is negative.

There are two general types of algorithms. In \textit{adaptive}
group testing, at each step the algorithm decides which group to test by observing
the responses of the previous tests. In \textit{non-adaptive} algorithm, all tests are carried out in parallel. There is a compromise algorithm between these two types, which is called a \textit{multistage} algorithm. For the multistage algorithm all tests are divided into $p$ sequential stages. The tests inside the same stage are performed simultaneously. The tests of the next stages may depend on the responses of the previous. In this context, a non-adaptive group testing algorithm is reffered to as a  one stage algorithm.
\subsection{Previous results}
We refer the reader to the monograph  \cite{DH} for a survey on group testing and its applications. In spite the fact that the problem of estimating the minimum \textit{average} (the set of defects is chosen randomly) number of tests has been investigated in many papers (for instance, see \cite{ds13, mt11}), in the given paper we concentrate our attention only on the minimal number of test in the \textit{worst case}. 

In 1982 \cite{dr82}, Dyachkov and Rykov proved that at least 
$$
\frac{s^2}{2\log_2(e(s+1)/2)}\log_2 t(1+o(1))
$$ 
%number of 
tests are needed for non-adaptive group testing algorithm. Recently, we have shown \cite{DVPS} that for non-adaptive group testing  $$\frac{s^2}{4e^{-2}\log_2s}\log_2 t(1+o(1))$$ tests are sufficient as $s\to\infty$. This result was obtained as the particular case of a more general bound for cover-free codes. 

 If the number of stages is $2$, then it was proved that $O(s \log_2 t)$ tests are already sufficient. It was shown by studying random coding bound for disjunctive list-decoding codes \cite{r90,d03} and  selectors \cite{BGV}. The recent work \cite{DVPS} has significantly improved the constant factor in the main term of number of tests for two stage group testing procedures. In particular, if $s\to\infty$, then 
 $$\frac{se}{\log_2e}\log_2t (1+o(1))$$
tests are enough for two stage group testing.  

As for adaptive strategies, there exist such ones that attain the information theory lower bound $s \log_2t (1+o(1))$. However, the number of stages in well-known optimal strategies  is a function of $t$, and grows to infinity as $t\to\infty$.

\subsection{Summary of the results}
In the given article we present some explicit algorithms, in which we make a restriction on the number of stages. It will be a function of $s$. We briefly give necessary notations in section \ref{Pre}. Then, in section~\ref{Hyp}, we present a general idea of searching defects using a hypergraph approach.  In section~\ref{Search2}, we describe a $4$-stage group testing  strategy, which detects $2$ defects and  uses the asymptotically optimal number of tests  $2\log_2t(1+o(1))$. As far as we know the best result for such a problem was obtained \cite{DSW} by Damashke et al. in 2013. They provide an exact two stage group testing strategy and use $2.5\log_2t$ tests. For other constructions for the case of $2$ defects, we refer to \cite{mr98, DL}.  In section~\ref{Searchs}, we propose searching of $s$ defects in $2s-1$ rounds. There we use $(2s-1)\log_2t(1+o(1))$ tests in the worst case. Finally, in Sect. \ref{FinT} for certain values of $t$ we provide the minimal number of tests of algorithm discussed in Sect. \ref{Search2}.
\section{Preliminaries}\label{Pre}

Throughout the paper we use $t$, $s$, $p$ for the number of elements, defectives, and stages, respectively. 
Let $\eq$ denote the equality by definition, $|A|$ -- the cardinality of the set $A$. The binary entropy function $h(x)$ is defined as usual $$h(x)=-x\log_2(x)-(1-x)\log_2(1-x).$$

A binary $(N \times t)$-matrix with $N$ rows $\x_1, \dots, \x_N$ and $t$ columns $\x(1), \dots, \x(t)$ (codewords)
$$
X = \| x_i(j) \|, \quad x_i(j) = 0, 1, \quad i \in [N],\,j \in [t]
$$
is called a {\em binary code of length $N$  and size $t$}.
The number of $1$'s in the codeword $x(j)$, i.e., $|\x(j)| \eq \sum\limits_{i = 1}^N \, x_i(j)= wN$,
is called the {\em weight} of $\x(j)$, $j \in [t]$ and parameter $w$, $0<w<1$, is the \textit{relative weight}.

One can see that the binary code $X$ can be associated with $N$ tests. A column $\x(j)$ corresponds to the $j$-th sample; a row $\x_i$ corresponds to the $i$-th test. 
Let $\u \bigvee \v$ denote the disjunctive sum of binary columns $\u, \v \in \{0, 1\}^N$.
For any subset $\S\subset[t]$ define the binary vector $$r(X,\S) = \bigvee\limits_{j\in\S}\x(j),$$
which later will be called the \textit{outcome vector}.

 By $\S_{un}$, $|\S_{un}|\le s$, denote an unknown set of defects. Suppose there is a $p$-stage group testing strategy $\mathfrak{S}$ which finds up to $s$ defects. It means that for any $\S_{un}\subset[t]$, $|\S_{un}|\le s$, according to $\mathfrak{S}$:
\begin{enumerate}
\item we are given with a code $X_1$ assigned for the first stage of group testing;
\item we can design a code $X_{i+1}$ for the $i$-th stage of group testing, based on  the outcome vectors of the previous stages $r(X_1,\S_{un})$, $r(X_2,\S_{un})$, \ldots, $r(X_i,\S_{un})$;
\item we can identify all defects $\S_{un}$ using $r(X_1,\S_{un})$, $r(X_2,\S_{un})$, \ldots, $r(X_p,\S_{un})$.
\end {enumerate}
 Let $N_i$ be the number of test used on the $i$-th stage and $$N_T(\mathfrak{S})=\sum_{i=1}^p N_i$$ be the maximal total number of tests used for the strategy $\mathfrak{S}$.
We define $N_p(t, s)$ to be the minimal worst-case total number of tests needed for group
testing for $t$ elements, up to $s$ defectives, and at most $p$ stages.

\section{Hypergraph approach to searching defects}\label{Hyp}
Let us introduce a hypergraph approach to searching defects. Suppose a set of vertices $V$ is associated with the set of samples $[t]$, i.e. $V = \{1,2\ldots, t\}$. 

\textbf{First stage:}
Let $X_1$ be the code corresponding to the first stage of group testing. For the outcome vector $r=r(X_1,\S_{un})$ let $E(r,s)$ be the set of subsets of $\S\subset V$ of size at most $s$ such that $r(X,\S)=r(X,\S_{un})$. So, the pair $(V,E(r,s))$ forms the hypergraph $H$. We will call two vertices \textit{adjacent} if they are included in some hyperedge of $H$. Suppose there exist a \textit{good} vertex coloring of $H$ in $k$ colours, i.e., assignment of colours to vertices of $H$ such that no two adjacent vertices share the same colour. By $V_i\subset V$, $1\le i\le k$, denote vertices corresponding to the $i$-th colour. One can see that all these sets are pairwise disjoint. 

\textbf{Second stage:}

Now we can perform $k$ tests to check which of monochromatic sets $V_i$ contain a defect. Here we find the cardinality of set $\S_{un}$ and $|\S_{un}|$ sets $\{V_{i_1},\ldots,V_{i_{|\S_{un}|}} \}$, each of which contains exactly one defective element.

\textbf{Third stage:}

Carrying out $\lev\log_2|V_{i_1}|\riv$  tests we can find a vertex $v$, corresponding to the defect, in the suspicious set $V_{i_1}$. Observe that actually by performing $\sum\limits_{j=1}^{\S_{un}}\lev\log_2|V_{i_j}|\riv$ tests we could identify all defects $\S_{un}$ on this stage.

\textbf{Fourth stage:}

Consider all hyperedges $e\in E(r,s)$, such that $e$ contains the found vertex $v$ and consists of vertices of $v\cup V_{i_2}\cup \ldots\cup V_{i_{|\S_{un}|}}$. At this stage we know that the unknown set of defects coincides with one of this hyperedges. To check if the hyperedge $e$ is the set of defects we need to test the set $[t]\backslash e$. Hence, the number of test at fourth stage is equal to degree of the vertex $v$.
%Finally, we have to find the remaining  $|\S_{un}|-1$ defects in the union of all such hyperedges, from which excludes the found vertex. 

%Note that if someone will prove the existence of a code $X_1$ such that maximal degree of vertices of hypergraph $H$ is upper bounded by $d$,
\section{Optimal searching of 2 defects}\label{Search2}
Now we consider a specific construction of $4$-stage group testing. Then we upper bound number of tests $N_i$ at each stage. 

\textbf{First stage:}

Let $C=\{0,1,\dots q-1\}^{\hat{N}}$  be the $q$-ary code, consisting of all $q$-ary words of length $\hat{N}$ and having size $t=q^{\hat{N}}$. 
Let $D$ be the set of all  binary words with length $N'$ such that the weight of each codeword is fixed and equals $ wN'$, $0<w<1$, and the size of $D$  is at least $q$, i.e., $q \le {N' \choose {wN'}}$.
On the first stage we use the concatenated binary code $X_1$ of length $N_1=\hat{N}\cdot N'$ and size $t=q^{\hat{N}}$, where the inner code is $D$, and the outer code is $C$. We will say $X_1$ consists of $\hat{N}$ layers.
Observe that we can split up the outcome vector $r(X_1,\S_{un})$ into $\hat{N}$ subvectors of lengths $N'$. So let $r_j(X_1,\S_{un})$ correspond to $r(X_1,\S_{un})$ restricted to the $j$-th layer. Let $w_j$, $j\in[\hat{N}]$, be the relative weight of $r_j(X_1,\S_{un})$, i.e., $|r_j(X_1,\S_{un})| = w_jN'$ is the weight of the $j$-th subvector of $r(X_1,\S_{un})$.

If $w_j=w$ for all $j\in[\hat{N}]$, then we can say that $\S_{un}$ consists of 1 element and easily find it.

If there are at least two defects, then  suppose for simplicity that $\S_{un}=\{1,2\}$. The two corresponding codewords of $C$ are $c_1$ and $c_2$. There exists a coordinate $i, 1\le i\le \hat{N}$, in which they differs, i.e., $c_1(i)\neq c_2(i)$.  Notice that the relative weight $w_i$ is bigger than $w$. 
%One can see that $w_j\ge w$ for any $j\in[\hat{N}]$. 
For any $i\in[\hat{N}]$ such that $w_i>w$, we can colour all vertices $V$ in $q$ colours, where the colour of $j$-th vertex is determined by the 
%restriction
corresponding $q$-nary symbol $c_i(j)$ of code $C$.
%of $\x(j)$  on the $i$-th layer. 
One can check that such a coloring is a good vertex coloring. 

\textbf{Second stage:}

We perform $q$ tests to find which coloured group contain $1$ defect.

\textbf{Third stage:}

Let us upper bound the size $\hat{t}$ of one of such suspicious group: 
$$\hat{t}\le {w_1N' \choose wN'}\cdot\ldots \cdot {w_{\hat{N}}N' \choose wN'}.$$
In order to find one defect in the group we may perform $\lev\log_2 \hat{t}\riv$ tests.

\textbf{Fourth stage:}

On the final step, we have to bound the degree of the found vertex $\it{v}\in V$ in the graph. The degree $\deg(\it{v})$ is bounded as
$$
\deg(\it{v})\le {wN' \choose (2w-w_1)N'}\cdot\ldots \cdot {wN' \choose (2w-w_{\hat{N}})N'}.
$$
We know that the second defect corresponds to one of the adjacent to $v$ vertices.
Therefore, to identify it we have to make $\lev\log_2\deg(\it{v})\riv$ tests.

Letting $\hat{N}$ tends to infinity we obtain the following bound:
\begin{align*}
\frac{N_T}{\log_2t}&\leq 
% \frac
% {{\hat{N}}\cdot N' + q + \max\limits_{w_i}\left(\lev\log_2\hat{t}\riv + \lev\log_2\deg(\it{v})\riv\right)}{\log_2t}
% \sim\\
% &\sim
\frac{
{{\hat{N}}\cdot N' + \max\limits_{w_i}\left(\log_2\hat{t} + \log_2\deg(\it{v})\right)}}
{(1+o(1))\hat{N}\log_2{N' \choose wN'}}.
\end{align*}
It is easy to see that in the worst case all values of $w_i$ are the same, hence
\begin{align}\label{finite_q}
\frac{N_T}{\log_2t}&\leq 
% \frac
% {{\hat{N}}\cdot N' + q + \max\limits_{w_i}\left(\lev\log_2\hat{t}\riv + \lev\log_2\deg(\it{v})\riv\right)}{\log_2t}
% \sim\\
% &\sim
\frac{
{{\hat{N}}\cdot N' + \max\limits_{w'}\log_2\left({w'N' \choose wN'}{wN' \choose (2w-w')N'}\right)}}
{(1+o(1))\hat{N}\log_2{N' \choose wN'}}.
\end{align}
By choosing the optimal parameter $w$, $wN'\in \mathcal{Z}$, we can minimize the number of tests for fixed value of $q$.

If $q\to \infty$, then we can rewrite \req{finite_q} as follows
%the total number of tests $N_T$  for the given $4$-stage searching procedure in the following manner:
$$
\frac{N_T}{\log_2t}\le \sup\limits_{w\le w'\leq \min(1, 2w)} f(w, w')(1+o(1)),
$$ where
\begin{align*}
f(w, w') &=
%\bigg({\hat{N}}\cdot N' + q + 
%{\hat{N}}\cdot w'N'\cdot h\left(\frac{w}{w'}\right)+\\
%&+{\hat{N}}\cdot wN'\cdot h\left(\frac{2w-w'}{w}\right)\bigg) / \log_2t=\\
%&=
\frac{1 + w'\cdot h\left(\frac{w}{w'}\right) + w\cdot h\left(\frac{2w-w'}{w}\right)}{h(w)}.
\end{align*}
Finally, we obtain the following 
%optimization problem
bound
%for continuous $w$ and $w'$
\begin{equation}\label{final_tests}
\frac{N_T}{\log_2t} \le \inf\limits_{0<w<1}\sup\limits_{w\le w'\le \min(1, 2w)} f(w, w').
\end{equation}
Let us find extreme value on $y$ of
$$
g(x,y) = y\cdot h(x/y) + x\cdot h((2x-y)/x).
$$ 
\begin{align*}
\frac{dg(x,y)}{dy}&=h(x/y)-\frac{x}{y}h'(x/y)-h'((2x-y)/x)=\\
&=
%-\frac{x}{y}\log_2\left(\frac{x}{y}\right)-\frac{y-x}{y}\log_2\left(\frac{y-x}{y}\right)-\frac{x}{y}\log\left(\frac{y-x}{x}\right)-\log_2\left(\frac{y-x}{2x-y}\right)
\log_2 y - 2\log_2(y-x)+\log_2(2x-y).
\end{align*}
This implies
$$
(y-x)^2 - 2xy +y^2=0.
$$
Hence, if we take $w=1/(2+\sqrt{2})$, then the supremum in (\ref{final_tests}) is attained at  $w'=1/2$, and achievable number of tests by $4$-stage group testing procedure is $2\log_2t(1+o(1))$. 

Observe that for fixed $q$ we can obtain only finite amount of rational values for parameter $w$, we could not provide an explicit construction of searching procedure with $2\log_2t(1+o(1))$ tests. But if $q\to \infty$, then the minimal number of test $N_T$ tends to $2\log_2t(1+o(1))$.

\section{Searching of $s$ defects}\label{Searchs}
Here we will use combination of the first two stages of the previous algorithm. Suppose the number of defects is at most $s$. In fact, we don't use  this fact in our algorithm. Let $C=\{0,1,\dots q-1\}^{{\hat{N}}}$, $|C|={q^{\hat{N}}}$, be the set of all $q$-ary words of length ${\hat{N}}$. Let $D$ be the set of all binary words of length $N'$ such that the weight of each codeword is fixed and equals $ N'/2 $, and the size of $|D|$ is at least $q$. On the first stage we use the concatenated binary code $X$ of length ${\hat{N}}\cdot N'$ and size $q^{\hat{N}}$, where the inner code is $D$, and the outer code is $C$. Notice that if the number of defects is one, then we are assumed to identify defect basing on the outcome vector $r_1(X,\S_{un})$. If this number is at least two than there exists a coordinate $i$ in which the corresponding $q$-ary words differs. It means that the outcome vector restricted on the $i$-th coordinate has the relative weight bigger than $w$. Split up all vertices $V$ in $q$ groups according to $q$-ary symbol in the $i$-th coordinate. On the next stage we perform $q$ tests and find which groups contain at least one defect. Then we will deal with each such group separately. If we could not divide a group into subgroups, then we easily find the unique defect in this group. In the worst case scenario, we have to perform $2s-1$ group testing stages, and the total number $N_T$ of tests is upper bounded by the sum of number of tests, which served for separating defects into disjoint groups, and number of tests, which used for finding $1$ defect among different groups. Thus, we have
$$
N_T\le (s-1){\hat{N}}\cdot N'+ s {\hat{N}}\cdot N'+q(s-1).
$$
In asymptotic regime, the total number of tests
$$
N_T \le (2s-1)\log_2 t(1+o(1)).
$$

\section{Finite number of objects}\label{FinT}
In this section we apply our $4$ stage procedure from~\ref{Search2} to specific values of $t$. Let us bound numbers of tests at each stage more properly. Recall that number of tests at the first stage $N_1$ is equal to $\hat{N}\cdot N'$.  In case $|S_{un}|=1$ we can find defective element based only on the outcome of the first stage of group testing.

Let $W=wN'$ and $W_i=w_iN'$. If our coloring is determined by symbols from $i$-th layer of the code $X_1$, then the actual number of suspicious sets equals $W_i \choose W$. Since we know exact number of defects it is sufficient to use ${W_i \choose W} - 1$ tests. Also note that we need to determine only one set with a defective element, therefore we can make ${W_i \choose W} - 2$ tests at the second stage.

The total number of elements in all suspicious groups is equal to
$$
{W_1 \choose W}\cdot\ldots \cdot {W_{\hat{N}} \choose W}.
$$
One can see that the numbers of elements of each color are the same, hence the cardinality $\hat{t}$ of one suspicious set is equal to
$$
\hat{t}=
{W_1 \choose W}\cdot\ldots \cdot {W_{\hat{N}} \choose W} / {W_i \choose W}
%\max\limits_{w\leq w_1 \leq \min(1, 2w)}{w_1N' \choose wN'}^{N_1 - 1}.
$$
So, at the third stage we need to perform $\lev\log_2\hat{t}\riv$ tests.
Before the last stage we have already known one of the defects. At each layer $j\ne i$ we have $W \choose 2W - W_j$ ways to choose $q$-nary coordinate of the second defect, but at the $i$-th layer we have only 2 suspicious coordinates left in the worst case. Therefore, the number of tests at the fourth stage is at most
$$
\lev \log_2\left(2
\frac{{W \choose 2W-W_1}\cdot\ldots \cdot {W \choose 2W-W_{\hat{N}}}}
{{W \choose 2W-W_i}}\right)\riv.
$$

We provide three tables with optimal values of tests for small $t \le 1000$, for $t=10^k$, $3 \le k\le 18$, and for some values of $t$, for which we have a small ratio of number of tests to $\log_2t$.
\begin{table}[ht]
		\caption{Number of tests for $t\leq 1000$} 
        \label{Number of tests for small values of $t$}
		\begin{center}
		\begin{tabular}{|c|c||c|c||c|c|}
			\hline
			$t$ & tests & $t$ & tests & $t$ & tests  \cr
            \hline
            & &  & & & \cr
			8-9 & $8$ &  29-36 & 14 & 126-256 & 20\cr
            10-16 & $10$ &  37-64 & 15 & 257-441 & 22\cr
            17-27 & $12$ &  65-81 & 16 & 442-784 & 24\cr
			28 & $13$ &  82-125 & 18 & 785-1000 & 25\cr
            \hline
		\end{tabular}
        \end{center}
\end{table}

In table \ref{Number of tests for $t=10^k$} and table \ref{Number of tests for $t$ with good ratio} we also present information bound $\underline{N}$, which is the minimum integer such that
$$
2^{\underline{N}}\geq 1 + {t \choose 1} + {t \choose 2}.
$$
 
\begin{table}[ht]
		\caption{Number of tests for $t=10^k$} 
        \label{Number of tests for $t=10^k$}
		\begin{center}
		\begin{tabular}{|c|c|c|c|}
        	\hline
            & & information & \cr
			$t=q^{N_1}$ & tests &  bound &  tests $/\log_2t$ \cr
            \hline
            & &  & \cr
			$10^3$ & 26 & 19 & $2.609$\cr
            $10^4$ & 33 & 26 & $2.483$\cr
            $10^5$ & 41 & 33 & $2.468$\cr
            $10^6$ & 48 & 39 & $2.408$\cr
            $10^7$ & 56 & 46 & $2.408$\cr
            $10^8$ & 64 & 53 & $2.408$\cr
            $10^9$ & 71 & 59 & $2.375$\cr
            $10^{10}$ & 79 & 66 & $2.378$\cr
            $10^{11}$ & 86 & 73 & $2.354$\cr
            $10^{12}$ & 94 & 79 & $2.358$\cr
            $10^{13}$ & 102 & 86 & $2.362$\cr
            $10^{14}$ & 109 & 93 & $2.344$\cr
            $10^{15}$ & 117 & 99 & $2.348$\cr
            $10^{16}$ & 124 & 106 & $2.333$\cr
            $10^{17}$ & 132 & 112 & $2.337$\cr
            $10^{18}$ & 139 & 119 & $2.325$\cr
            \hline
		\end{tabular}
        \end{center}
\end{table}

\begin{table}[ht]
 		\caption{Number of tests for $t$ with small ratio $\text{tests }/\log_2t$}
        \label{Number of tests for $t$ with good ratio}% $\text{tests }/\log_2t$}
  		\begin{center}
  		\begin{tabular}{|c|c|c|c|}
  			\hline
            & & information & \cr
			$q^{N_1}=t$ & tests &  bound &  tests $/\log_2t$ \cr
            \hline
            & &  & \cr
  			$28^2=784$ & 24 & 19 & $2.496$\cr
            $15^3=3375$ & 29 & 23 & $2.474$\cr
            $21^3=9261$ & 32 & 26 & $2.428$\cr
            $28^3=21952$ & 35 & 28 & $2.427$\cr
            $15^4=50625$ & 37 & 31 & $2.368$\cr
            $21^4=194481$ & 41 & 35 & $2.334$\cr
            $21^5=4084101$ & 51 & 43 & $2.322$\cr
            $15^6=11390625$ & 54 & 46 & $2.304$\cr
            $21^6=85766121$ & 60 & 52 & $2.277$\cr
            $21^9=794280046581$ & 89 & 79 & $2.251$\cr
            %$350277500542221=21^{11}$ & 108 & 96 & $2.235$\cr   
            $21^{11}\approx  3.5\cdot 10^{14}$ & 108 & 96 & $2.235$\cr   
            \hline
 		\end{tabular}
        \end{center}
\end{table}

% Summing up, the total number of tests is upper bounded by
% $$
% \hat{N}\cdot N' + {w_iN' \choose wN'} - 2 + max\
% $$

\textbf{Acknowledgements.}

I.V. Vorobyev, N.A. Polyanskii and V.Yu. Shchukin have been supported in part by the Russian Science Foundation
under Grant No. 14-50-00150.

\end{document}